\documentclass[twocolumn]{aastex631}

\usepackage{amsmath}	
\usepackage{amssymb}
\usepackage{bm}	
\usepackage{lipsum}
\usepackage{afterpage}
\usepackage{multirow}
\usepackage{makecell}
\usepackage[figuresright]{rotating}
\usepackage{tabularx}

\newcommand{\src}{MAXI~J1803$-$298}
\newcommand{\nicer}{\textit{NICER}}

\begin{document}

\title{Variable ionized disk winds in \src\ revealed by \nicer}

\correspondingauthor{Cosimo Bambi} 
\email{bambi@fudan.edu.cn}

\author[0000-0003-0847-1299]{Zuobin Zhang}
\affiliation{Center for Astronomy and Astrophysics, Center for Field Theory and Particle Physics and Department of Physics, Fudan University, Shanghai 200438, China}

\author[0000-0002-3180-9502]{Cosimo Bambi}
\affiliation{Center for Astronomy and Astrophysics, Center for Field Theory and Particle Physics and Department of Physics, Fudan University, Shanghai 200438, China}
\affiliation{School of Natural Sciences and Humanities, New Uzbekistan University, Tashkent 100007, Uzbekistan}

\author{Honghui Liu}
\affiliation{Center for Astronomy and Astrophysics, Center for Field Theory and Particle Physics and Department of Physics, Fudan University, Shanghai 200438, China}
\affiliation{Institut f\"ur Astronomie und Astrophysik, Eberhard-Karls Universit\"at T\"ubingen, D-72076 T\"ubingen, Germany}

\author[0000-0002-9639-4352]{Jiachen Jiang}
\affiliation{Institute of Astronomy, University of Cambridge, Madingley Road, Cambridge CB3 0HA, UK}
\affiliation{Department of Physics, University of Warwick, Gibbet Hill Road, Coventry CV4 7AL, UK}

\author{Fangzheng Shi}
\affiliation{Shanghai Astronomical Observatory, Chinese Academy of Sciences, 80 Nandan Road, Shanghai 200030, China}

\author{Yuexin Zhang}
\affiliation{Center for Astrophysics, Harvard \& Smithsonian, 60 Garden St, Cambridge, MA 02138, USA}
\affiliation{Kapteyn Astronomical Institute, University of Groningen, P.O.\ BOX 800, 9700 AV Groningen, The Netherlands}

\author{Andrew J. Young}
\affiliation{School of Physics, Tyndall Avenue, University of Bristol, Bristol BS8 1TH, UK}

\author{John A. Tomsick}
\affiliation{Space Sciences Laboratory, 7 Gauss Way, University of California, Berkeley, CA 94720-7450, USA}

\author{Benjamin M. Coughenour}
\affiliation{Space Sciences Laboratory, 7 Gauss Way, University of California, Berkeley, CA 94720-7450, USA}

\author{Menglei Zhou}
\affiliation{Institut f\"ur Astronomie und Astrophysik, Eberhard-Karls Universit\"at T\"ubingen, D-72076 T\"ubingen, Germany}

\begin{abstract}

We present the results from the \nicer\ observation data of \src\ across the entire 2021 outburst. In the intermediate and soft state, we detect significant absorption lines at $\sim 7.0$~keV and $\sim 6.7$~keV, arising from X-ray disk winds outflowing with a velocity of hundreds of km s$^{-1}$ along our line of sight. The fitting results from photoionized model suggest that the winds are driven by thermal pressure and the mass-loss rate is low. We find a clear transition for iron from predominantly H-like to predominantly He-like during the intermediate-to-soft state transition. Our results indicate this transition for iron is caused by the evolution of the illuminating spectrum and the slow change of the geometric properties of the disk winds together. The coexistence of disk winds and QPOs features in intermediate state is also reported. Our study makes \src\ the first source in which a transition from optical winds to X-ray winds is detected, offering new insights into the evolution of disk winds across an entire outburst and long-term coupling of accretion disks and mass outflows around accreting black holes.


\end{abstract}


\section{Introduction}

X-ray disk winds are one of the important forms of outflows, frequently observed in X-ray binaries and active galactic nuclei \citep{Ponti2012, Parker2017}. Appearing in the form of outflowing hot ionized gas emitted from the accretion disk, X-ray disk winds are more prevalently detected in high-inclination systems, as illustrated in \cite{Ponti2012}, which indicate a biconical, roughly equatorial geometry and an origin from the accretion disk. They are typically highly ionized, such that hydrogen- and helium-like iron tend to dominate the spectra with strong absorption lines at $\sim7$~keV and 6.7~keV, respectively.

The significant role of X-ray disk winds was already evident from theoretical studies (e.g., \citealt{Begelman1983}), and it was quickly confirmed with high-resolution X-ray observations. For instance, observations suggest the mass outflow rates carried away by X-ray disk winds can be comparable or even significantly larger than the mass accretion rates (e.g., \citealt{Ueda2009}; \citealt{Ponti2012}). Observations show anti-correlation between the winds and jets (e.g., \citealt{Neilsen2009}), which indicate the winds, as alternative mass ejection modes, exert significant influence on the formation of jets or even quenches the jets.

The physical mechanism to produce X-ray disk winds is not unambiguously established yet. There are at least three possible mechanisms to drive them: thermal pressure (e.g., \citealt{Done2018}), magnetic pressure (e.g., \citealt{Fukumura2017}), or radiation pressure (e.g., \citealt{Higginbottom2015}). However, it is difficult to distinguish different launching mechanisms due to the uncertainty in measurement of the density and radial location of the disk winds. Moreover, it is also possible that different mechanisms may drive the disk winds at the same time (e.g., \citealt{Neilsen2012}). 


X-ray disk winds of black hole X-ray binary system appear to be state–dependent. The outburst of a black hole X-ray binary system is characterized by the changes in the relative contribution of thermal and non-thermal emission components occurring throughout the outburst, resulting in distinctive spectral states (e.g., \citealt{Remillard2006}). During the early stage of the outburst, the source is in a low hard state, with the X-ray spectrum dominated by non-thermal emission in the form of a hard power-law. The spectrum remains hard as the source brightens, and eventually transitions into a thermal emission-dominated soft state over the course of a few days. During this transition, the contribution of thermal emission to the spectrum gradually increases, and the source passes through the intermediate state. X-ray disk winds are detected in spectrally–soft, disk–dominated states, but not clearly detected in the low/hard state (e.g., \citealt{Miller2006}; \citealt{Miller2008}; \citealt{Neilsen2009}; \citealt{Ponti2012}).

On the other hand, low-ionization (cold) disk winds have been detected via optical/infrared observations in some sources (e.g., V404~Cyg, \citealt{Munoz2016}; MAXI~J1820+070, \citealt{Munoz2019}). Observations show that optical winds are always detected during the hard state, extensively during outburst rise, but also during the decay toward the end of the soft-to-hard transition \citep{Munoz2019}, while infrared winds are ubiquitously detected across the entire outburst \citep{Sanchez2020}. Given that X-ray disk winds are typically observed during soft but not hard states, an evolution of the visibility of the winds across the outburst is expected \citep{Sanchez2020}. In addition, \citet{Castro2022} reported the transient neutron stars binary Swift J1858.6-0814 exhibits wind-formed, moderately ionized and blue-shifted absorption in time-resolved UV spectroscopy during a luminous hard state.


\src\ was first discovered in the early stages of an outburst on May 1 2021 by the Monitor of All-sky X-ray Image (\textit{MAXI}; \citealt{Serino2021}). Both \nicer\ (\citealt{Gendreau2021}) and \textit{Swift} quickly localized the event, with the latter also detecting an optical counterpart \citep{Gropp2021}. Following the discovery, many follow-up observations were performed in X-ray, optical and radio wavelengths. Optical spectroscopy, using the Southern African Large Telescope, indicated that the source is likely a system known as a low-mass black hole X-ray binary \citep{Buckley2021}, which was confirmed by \textit{MAXI/ GSC} observation to the hard-to-soft transition of the source. Additionally, \cite{Espinasse2021} detected a radio counterpart of the source. Subsequent observations of the source in the X-ray band, performed with various telescopes, have revealed the detection of low-frequency quasi-periodic oscillations (LFQPOs) (\citealt{Chand2022}; \citealt{Coughenour2023}; \citealt{Zhu2023}) and reflection features (\citealt{Bambi2021}; \citealt{Coughenour2023}; \citealt{Feng2022}). 

Both \textit{NuSTAR} and \textit{NICER} observations revealed periodic light curve dips (\citealt{Xu2021}; \citealt{Gendreau2021}). \citet{Jana2022} carefully studied the periodic dips on light curve with \textit{AstroSAT} and \textit{NuSTAR} observations, and argued that periodic dips appear with a periodicity of $7.02\pm0.18$~h and are caused by the obscured materials of the bulge (thickened material) of the outer accretion disc. That study strongly suggests a high inclination angle of approximately $70^\circ$ or above. The absorption-like feature probably arising from X-ray disk winds were first reported with \textit{Swift} observations obtained on 2021 May 20-21 when the source was in the intermediate state \citep{Miller2021}. Furthermore, optical spectroscopy conducted in the hard state detected signs of an outflow, as indicated by the presence of p Cygni-like profiles in hydrogen Balmer lines \citep{Buckley2021}. However, for \src, it is still unclear whether X-ray disk winds exist in various accretion states and how they evolve with the spectral characteristics of the source. In this paper, we analyze the observational data from the \nicer\ observation campaign, investigating the nature and evolution of the X-ray disk winds. 


The paper is organized as follows. In Sec.~\ref{observations}, we present the observational data reduction. The spectral analysis is reported in Sec.~\ref{analysis}. We discuss the results and report our conclusions in Sec.~\ref{discussion} and Sec.~\ref{conclusion}, respectively.

\section{Observation and data reduction} 
\label{observations}

Since the day after its discovery on 2021 May 1 (MJD 59335), \nicer\ began to perform extensive high-cadence monitoring of this source, covering the entire outburst.

\begin{figure}
    \centering
    \includegraphics[width=0.99\linewidth]{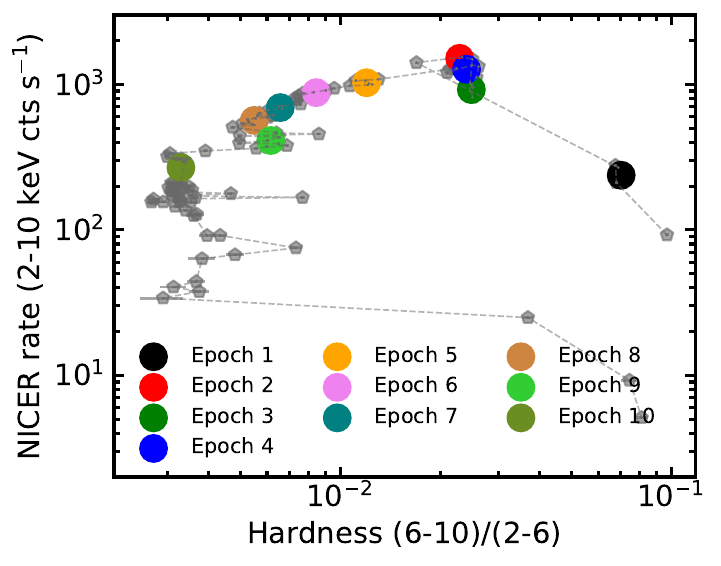} \\
    \caption{Hardness intensity diagram for the 2021 outburst of \src\ as seen by \nicer\ (gray pentagons). The hardness is defined as the ratio between count rates in the $6–10$~keV and $2–6$~keV bands. We use colorful points to mark the spectroscopic epochs analyzed in this draft. Ten epochs cover the transition process from the hard state to the soft state. Epoch~1 is identified as hard state, Epochs~2--4 as intermediate state, and Epochs~5--10 as soft state.}
    \label{nicer_hid}
\end{figure}

We process the \nicer\ data using \nicer\ data analysis software (NICERDAS) (version 2021-04-01\_V008) and CALDB version 20210707. We first conduct the standard \nicer\ reduction routine \texttt{nicerl2} to reduce the data with the default filtering criteria: the pointing offset is less than $54^{\prime \prime}$, and the pointing direction is more than $40^{\circ}$ away from the bright Earth limb, more than $30^{\circ}$ away from the dark Earth limb, and outside the South Atlantic Anomaly. The "undershoot" or "overshoot" (flagged as EVENT\_FLAGS = bxxxx00) and forced triggers (flagged as EVENT\_FLAGS=bx1x000) filtering are performed. In addition, we remove the data of detectors \# 14 and \# 34 because of electronic noise. The energy spectra of the background is extracted with the \texttt{nibackgen3C50} tool \citep{Remillard2022}. We create the Redistribution Matrix File (RMF) and Ancillary Response File (ARF) using the tasks \texttt{nicerl3}. Lightcurves with 1~s time resolution are produced from the full energy band, as well as the 2--6~keV band and the 6--10~keV band in order to calculate the hardness, which is defined as the ratio between the count rates in the $6-10$~keV and $2-6$~keV band. Figure~\ref{nicer_hid} shows the hardness intensity diagram (HID; \citealt{Homan2001}) with \nicer\ data. The resulting HID follows an anti-clockwise q-shaped hysteresis pattern, which is a behavior typically seen in black hole low-mass X-ray binaries (e.g., \citealt{Fender2004}). 


Unless stated otherwise, the uncertainties provided in this work are at the 90\% confidence limit. We perform all spectral fitting using XSPEC 12.11.1 \citep{Arnaud1996}. For all fits, we utilize the wilm set of abundances \citep{Wilms2000} and vern photoelectric cross sections \citep{Verner1996}. The $\chi^2$ fit statistics are employed for our analysis.


\begin{figure*}
    \centering
    \includegraphics[width=0.75\linewidth]{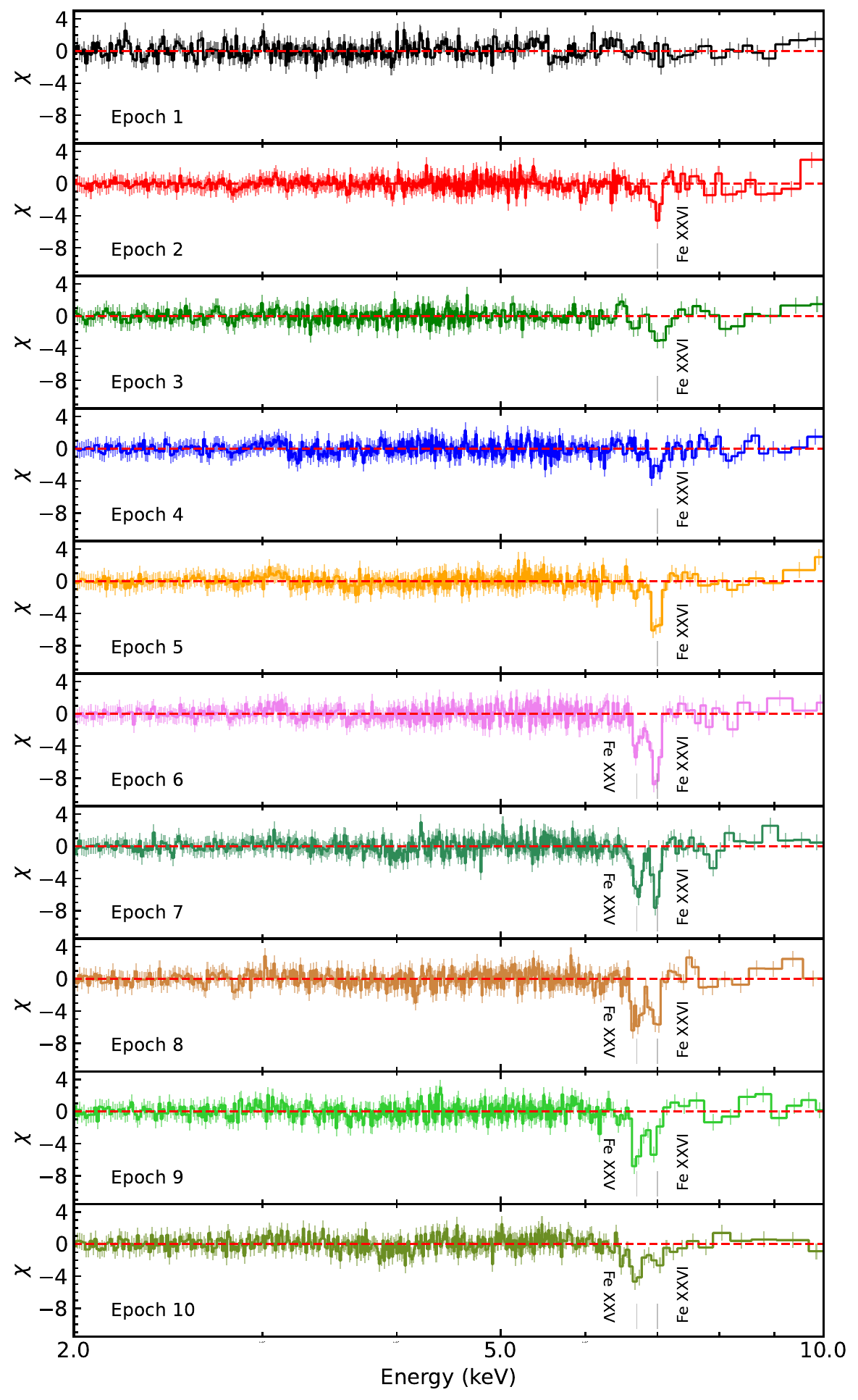} \\
    \caption{Residuals to the model: \texttt{Tbabs $\times$ (diskbb + nthcomp + gauss)} with three additional emission gaussian lines for the calibration issues in the $2.0-3.0$~keV band. Data are rebinned for visual clarity. The two vertical gray lines mark the positions of absorption by Fe XXV Ly$\alpha$ and Fe XXVI Ly$\alpha$ in the rest frame.}
    \label{ratio}
\end{figure*}


\begin{figure}
    \centering
    \includegraphics[width=0.99\linewidth]{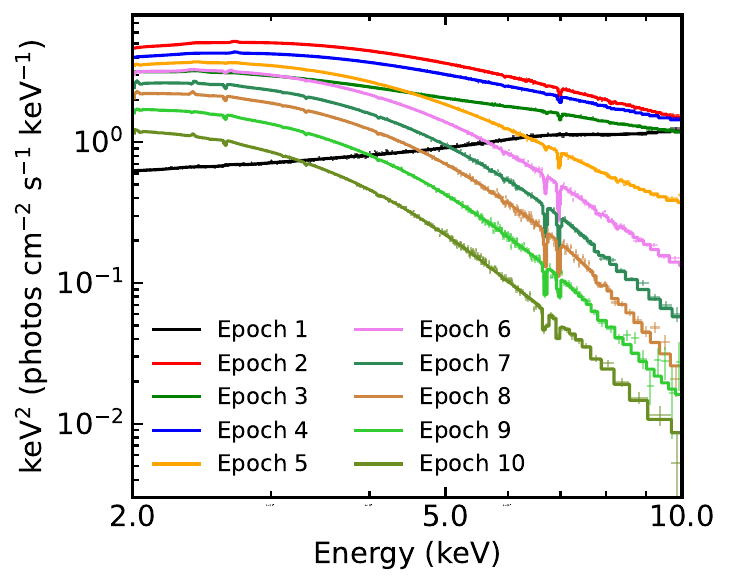} \\
    \caption{The unfolded spectra of Epoch~1--10. It is clearly shown that the disk wind features evolve with the continuum spectrum profile and flux. Data are rebinned for visual clarity.}
    \label{spectra}
\end{figure}

\section{Data analysis} \label{analysis}


For spectral analysis, we only consider the 2–10~keV band, ignoring the data below 2~keV because of large calibration uncertainties in the low energy band. A systematic error of 0.5\% is added to the \nicer\ spectra. We start the fitting with the following model combination: \texttt{Tbabs $\times$ (diskbb + nthcomp + gauss)}. \texttt{Tbabs} is incorporated to account for the absorption caused by the interstellar medium. \texttt{diskbb} \citep{Mitsuda1984} and \texttt{nthcomp} (\citealt{Zdziarski1996}; \citealt{Zycki1999}) account for the multi-temperature thermal disk blackbody emission from the disk and the inverse-Comptonized emission from the corona, respectively. $KT_{\rm e}$ (electron temperature) is fixed at $100$~keV. We link $KT_{\rm bb}$ (seed photon temperature) in \texttt{nthcomp} to $T_{\rm in}$ in \texttt{diskbb}. The source has been reported to have strong reflection features (\citealt{Feng2022}; \citealt{Coughenour2023}), and we use a simple \texttt{gauss} model to fit them. With this model combination, we observe notable residuals in the soft X-ray range, particularly three Gaussian-like narrow features around 2.0~keV, 2.4~keV, and 2.7~keV. These energy values correspond to specific features present in the effective area of \nicer\ versus energy, as the case in \cite{Wang2020}. Therefore, we include three \texttt{gauss} models for calibration issues and adopt the following model combination for the continuum fitting: \texttt{Tbabs $\times$ (diskbb + nthcomp + gauss + gauss + gauss + gauss)}. 


We go through all the observation data from \nicer\ and find that there are absorption-like features in the spectra caused by the X-ray disk winds, when the source is in a soft state. However, these weak features are not significant enough ($ > 3\sigma $) to be confirmed. Therefore, we merge nearby observations with similar hardness and flux levels in chronological order to obtain higher quality spectra with longer exposure time. It is worth noting that after July 27th, the source completely entered the soft state, with the spectrum being dominated by thermal component generated by the accretion disk. The background played a crucial role at the energy band of the potential absorption lines, making it extremely difficult to confirm the absorption lines caused by the X-ray disk winds. Thus, we do not include the observation after July 27th. During the transition from the soft state back to the hard state, the source was in a low-flux state, and \nicer\ did not conduct sufficiently intensive observations, preventing us from obtaining high-quality spectra. Therefore, we also excluded that period from this study. 

Finally, we obtained ten sets of observation data, which are labeled as Epoch~1 to Epoch~10. Figure~\ref{nicer_hid} illustrates the positions of these ten epochs in the HID. As shown in the figure, they precisely cover the transition process from the hard state to the soft state.

In order to highlight the absorption, we first fit each spectrum in the 2.0–10~keV energy range using the phenomenological model mentioned above: \texttt{Tbabs $\times$ (diskbb + nthcomp + gauss + gauss + gauss + gauss)}. Figure~\ref{ratio} shows the residuals of data to best-fit model for all ten spectra. Data are rebinned for illustrative purposes. We note the presence of features in the spectrum at $\sim 6.7$~keV and $\sim 7$~keV, typical of Fe XXV K$\alpha$ and/or Fe XXVI K$\alpha$ absorption line in Epoch~2-10. No absorption line is visible in Epoch~1.


Following on from this, we step a Gaussian line (varying $\sigma$ and the normalization, allowed to be positive or negative) for a blind line research over 6-10~keV, and record the $\Delta \chi^2$ at each point of this grid. Comparing the change in $\chi^2$, we can distinguish the significance of absorption lines in every epoch. With the threshold of $3\sigma$, common line for Epochs~2-10 is Fe XXVI K$\alpha$ at $\sim7$~keV. Fe XXV K$\alpha$ absorption line (at $\sim 6.7$~keV) is identified in Epochs~6-10. We have not detected any significant absorption lines in Epoch~1. The absorption lines that can be identified and have a significance of $>3 \sigma$ are all marked in Figure~\ref{ratio}.


For a more physical description, we calculate a grid model with the \texttt{XSTAR} code (\citealt{Kallman2001}; \citealt{Kallman2004}) for each of the epochs with significant absorption features. In these calculations, we assume the absorption corrected best-fit continuum model as the seed ionizing spectrum. The turbulent velocity is set to be 300~km~s$^{-1}$ and the density ($n$) to $10^{15}$~cm$^{-3}$. We fit three free parameters with \texttt{XSTAR} grid model, including the column density of the absorber ($N_{\rm H}$), ionization parameter ($\log \xi$), and redshift ($z$). Table~\ref{bestfit} shows the best-fit parameters obtained for Epochs~1-10.


\section{Discussion} \label{discussion}

Based on the results obtained from the continuum modelling, in combination with the positions on the HID diagram, we classify the epochs into accretion states as follows: Epoch~1 to hard state, Epochs~2-4 to intermediate state and Epochs~5-10 to soft state. We detect significant K$\alpha$ absorption lines from H-like (and He-like in Epochs~6--10) iron in the disk-dominated X-ray spectrum of \src\ during its transition from intermediate-to-soft state. The presence of an outflow is supported by the inferred blue-shift of the absorption. Given the high inclination angles of $\sim 70^{\circ}$ derived from the X-ray dips and reflection study (\citealt{Coughenour2023}; \citealt{Jana2022}), this source joins the growing number of high inclination X-ray binaries with such absorption features. 

\subsection{Nature of the accretion X-ray disk winds}

The disk winds can be driven by three primary mechanisms: thermal pressure \citep{Done2018}, radiation pressure \citep{Fukumura2017}, and magnetic pressure \citep{Higginbottom2015}. In the case of \src, we find high-ionization state disk winds, which can exclude the radiation pressure as the dominant mechanism since radiation pressure depends mostly on UV absorption lines and is only effective when $\log \xi <3$ \citep{Proga2002, Proga2003}. One crucial proxy that distinguishes the magnetically driven wind and thermally driven wind is the launching location of disk winds. The magnetically driven wind is always launched from the inner part of disk in the low Eddington ratio regime, while the thermally driven wind can only be launched from relatively large distances ($\sim 10^4$~$R_{\rm g}$) \citep{Begelman1983, Woods1996}, where $R_{\rm g} \equiv GM/c^2$ is the gravitational radius. 

The wind launching radius can be estimated from the parameters obtained with \texttt{XSTAR}. With the assumption that the outflow velocity of the wind is larger than the local escape velocity, we have the equation for the lower limit of the wind location, $r_{\rm min} = 2GM_{\rm BH}v_{\rm out}^{-2}$. The \nicer\ spectra are biased towards a low blue-shift absorption line ($z > -0.004$), which indicates disk winds with a velocity of $< 1200$~km/s. We adopt this value and get the lower limit of wind location $r_{\rm min} > 10^4$~$R_{\rm g}$. Such an estimation suggests that thermal pressure is the dominant process for the X-ray disk winds in \src, as was proposed in 4U~1630--47 \citep{Kubota2007}, H~1743--322 \citep{Miller2006} and GRS~1915+105 \citep{Ueda2009}. We note that the absolute energy scale is not known precisely, which means that the exact blueshift is difficult to determine.

The mass outflow rate carried away by the X-ray disk winds is evaluated as $$\Dot{M}_{\rm wind}=4\pi R^2nm_{\rm p}v_{\rm out}\frac{\Omega}{4\pi}=4\pi m_{\rm p}v_{\rm out}\frac{L_{\rm X}}{\xi}\frac{\Omega}{4\pi}$$ where $\Omega/4\pi$ is the covering factor, $m_{\rm p}$ is the proton mass, $v_{\rm out}$ is the wind outflow velocity. In this case, we assume a covering factor of $1/2$. A black hole mass of 8.0~$M_{\bigodot}$ and a distance of 10.0~kpc are considered in our calculation \citep{Mata2022, Shidatsu2022}. Assuming the parameters obtained with \texttt{XSTAR}, this gives $$\Dot{M}_{\rm wind}=0.2 \sim 5 \times 10^{18} \, \, {\rm g} \, \, {\rm s}^{-1}.$$ Meanwhile, the mass accretion rate in the inner part of the disk is estimated as $$\Dot{M}_{\rm acc}=\frac{L_{\rm X}}{\eta c^2}$$ where the efficiency is assumed to be $\eta=0.1$. Our analyses suggest $$\Dot{M}_{\rm wind}=0.01 \sim 0.1 \Dot{M}_{\rm acc}.$$


\cite{Ponti2012} compiled a sample of source with X-ray disk wind features observed with \textit{Chandra}, \textit{XMM–Newton} and \textit{Suzaku}, and sorted these sources on the $\Dot{M}_{\rm wind}/\Dot{M}_{\rm acc}$ versus luminosity plane. \src\ shows a luminosity of $\sim 0.03-0.09$~$L/L_{\rm Edd}$ in this state, and disk winds carrying away lower mass than the one accreted into the central object, which makes the source lying at the low luminosity part, and following the predicted line for thermally driven disk winds in Fig.~5 of \cite{Ponti2012}. Note that there are several uncertainties from the continuum modelling and the measurement of distance and mass, which may affect our calculations.





\begin{figure}
    \centering
    \includegraphics[width=0.99\linewidth]{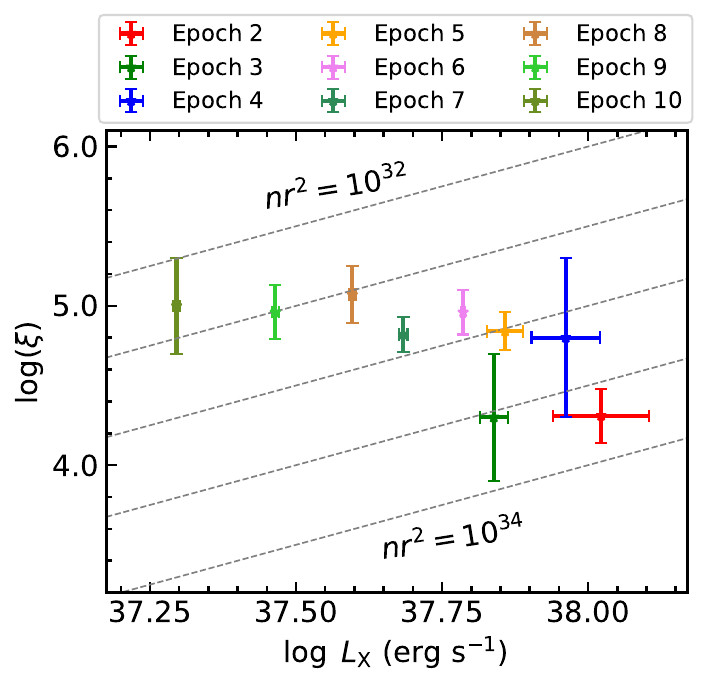} \\
    \caption{$\log \xi$ versus unabsorbed luminosity in the $1.0-10.0$~keV band. The parameters are obtained from the \texttt{XSTAR} fit for Epoch~2--10. A slowly change of physical properties of the X-ray wind is suggested during intermediate-to-soft state transitions.}
    \label{ion_luminosity}
\end{figure}


\begin{figure*}
    \centering
    \includegraphics[width=0.95\linewidth]{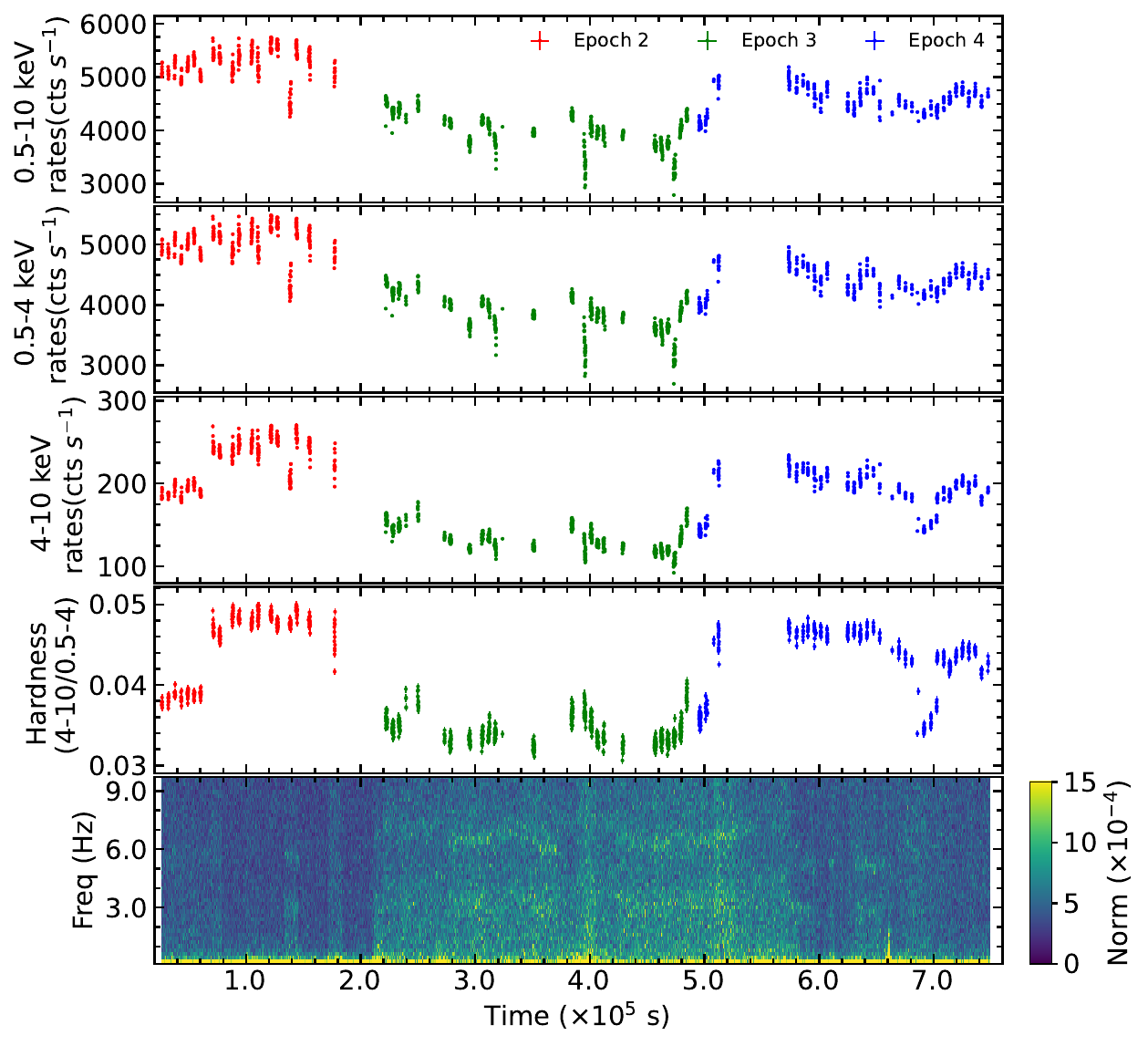} \\
    \caption{\nicer\ light curve with a time bin size of 30~s derived in the 0.5-10~keV (top panel), 0.5-4~keV (middle top panel), 4-10~keV (middle panel), and hardness derived between the 4–10 and 0.5–4~keV bands (middle bottom panel). Corresponding dynamical power density is shown in bottom panel.}
    \label{nicer_lcurve}
\end{figure*}


\begin{figure}
    \centering
    \includegraphics[width=0.99\linewidth]{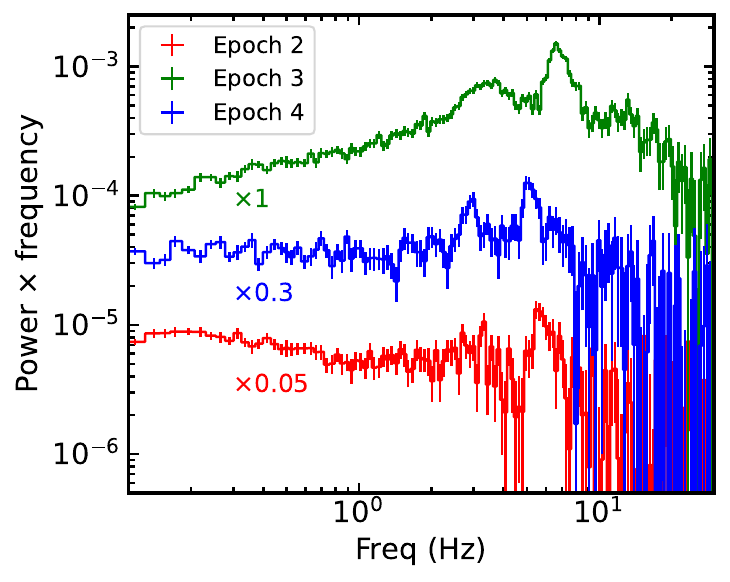} \\
    \caption{Power density spectra of Epoch~1 (red), Epoch~2 (green), and Epoch~3 (blue), scaled with frequency to illustrate the potential QPOs rather than the low-frequency red noise. We artificially multiply the power by a different factor for plotting clarity.}
    \label{pds}
\end{figure}


\begin{figure}
    \centering
    \includegraphics[width=0.99\linewidth]{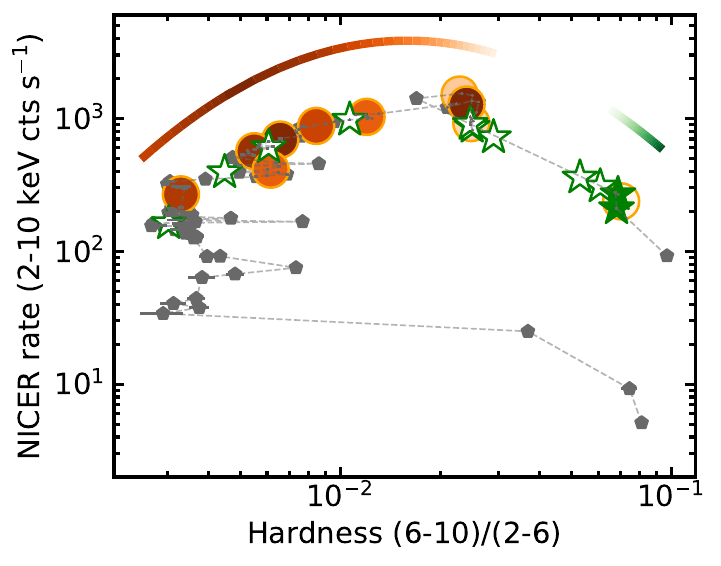} \\
    \caption{Sketch of the wind evolution across the hardness-intensity diagram of \src\ (based on \citealt{Mata2022} and this work). We use stars to mark the spectroscopic epochs of optical observations, with solid stars for the observation showing optical wind features, and hollow stars for the observations without optical wind features. We use circles to mark the ten epochs analyzed in this work. The darkness of the filling color represents the intensity of the X-ray disk winds (using the column density as proxy). The stronger the X-ray wind, the darker the filling color. The blurred green and red line indicate the (approximate) location of the wind detections at optical and X-ray band, respectively. The darkness of the color of red line show the strength of X-ray disk winds. A transition from optical wind to X-ray wind is clearly shown.}
    \label{wind_tran}
\end{figure}


\subsection{Evolution of the X-ray disk winds}

For \src, no X-ray disk wind feature is detected in Epoch~1, when the source is in the hard state, while we detect significant X-ray disk winds in the intermediate state and soft state. It is consistent with the conclusion that X-ray disk winds are mainly found in soft states of low mass X-ray binaries. Through the analyses described in Section~\ref{analysis}, we find a clear transition for iron from predominantly H-like to predominantly He-like during the intermediate state to soft state transition.

To reveal the physical process behind this evolution, we show in Figure~\ref{ion_luminosity} the correlation between the ionization of disk winds and the source X-ray luminosity. The ionization of disk winds is defined as $\xi=L_{\rm X}/nr^2$ (where $n$ and $r$ are the number density and the distance of the absorbing material from the illuminating source). If we assume that the geometric properties of the plasma do not change in the transition between different accretion states (i.e., $nr^2={\rm constant}$), a linear correlation is excepted for X-ray luminosity and ionization state of disk winds. We trace the path of the best-fit value and find the intermediate to soft state transitions are probably indicated by a slowly change of geometric properties ($nr^2$). In fact, we cannot rule out the possibility that geometric properties may keep stable during this process, because for many observations we can only get the lower limit of this parameter, and the values of $nr^2$ represented by different dotted lines in the Figure~\ref{ion_luminosity} do not change dramatically. In general, we are inclined to think the evolution of the illuminating spectrum and the slow change of the geometric properties of the disk winds together lead to the transition for iron from predominantly H-like to predominantly He-like.

\citet{Coughenour2023} and \citet{Zhu2023} reported the existence of QPOs in the intermediate state of the source. Combined with the characteristics of the disk winds detected in the intermediate state, it indicates that both QPOs and disk winds exist in this state, i.e., Epoch~2 to Epoch~4. For a more detailed timing analysis, we make light curves of the dispersed spectrum from each observation in the 0.5–10.0~keV, 0.5–4.0~keV and 4.0–10.0~keV band. The light curves are presented in Figure~\ref{nicer_lcurve}. The top panel shows that the observed 0.5–10~keV count rate of \src\ exhibited a low flux state by a factor of 1.4. Then the source returned to the original high flux level after 300~ks. The corresponding 4–10~keV count rate changed by a factor of 2.0, suggesting that the hard X-ray flux changed more significantly than the soft X- ray flux. The X-ray hardness of \src\ is redefined as the count rate ratio between the 4–10 and 0.5–4~keV bands. We notice the hardness was consistently higher in the high flux state than that in the low flux state, as shown in Figure~\ref{nicer_lcurve}. Timing analysis shows a transition of QPOs signatures during this flux drop event, as shown in the bottom panel in Figure~\ref{nicer_lcurve}. The time-averaged power density spectra (PDS) of Epoch~2, 3 and 4 are shown in Figure.~\ref{pds}.

Epochs~2--4 basically correspond to the high flux state before the flux drop, the low flux state, and the high flux state after the flux drop. Generally speaking, Epoch~3 shows stronger variability than Epoch~2 and Epoch~4. The PDS of the source are markedly different, but not entirely featureless over this flare event. The PDS of Epoch~4 show two potential QPO signals at 3.0 and 5.1~Hz. Since these two frequencies are not harmonically related, these signals must be entirely separate. The coexistence of a Type-B and C QPOs is preferred, as the \textit{NuSTAR} data shows \citep{Coughenour2023}. The identification QPO type is also different in Epoch~3, because there is a broad hump at $\sim3$~Hz, which is unusual in XRBs. The prompt flux variation detected along with the emergence of different QPO behaviors do not significantly affect the weakness of disk wind strength. The event is an interesting case for investigating the origin and structure of disk winds in black hole X-ray binaries. More detailed temporal-spectral analysis will be shown in a forthcoming paper.


\subsection{A transition from optical winds to X-ray disk winds}

Disk winds are detected in three energy band: X-ray, optical, and near-infrared (NIR) band. The X-ray disk winds is revealed by X-ray spectroscopy through the blue-shifted absorption lines of mildly to highly ionised species (e.g., \citealt{Miller2006}), and preferentially detected in the soft state of high inclination sources \citep{Neilsen2009, Ponti2012}. In the optical band, lower ionization wind is detected in optical emission lines of (mainly) helium and hydrogen series as P-Cyg profiles, and typically appears during the hard state (e.g., \citealt{Munoz2016, Munoz2019}; \citealt{Mata2018}). Unlike the optical wind signatures, the NIR features are not exclusive to the hard state and are also detected in the soft state. A sketch that broadly summarises the visibility of the wind across the outburst of typical X-ray binaries and as a function of the spectral band is presented in Fig.~4 in \citet{Sanchez2020}.

To date, up to six LMXBs have been found to exhibit P-Cygni profiles in their NIR and/or optical spectra: GX~13+1 \citep{Bandyopadhyay1997}, V404~Cyg (\citealt{Munoz2016}; \citealt{Munoz2017}; \citealt{Mata2018}), V4641~Sgr \citep{Munoz2018}, Swift~J1858.6--0814 \citep{Munoz2020}, MAXI~J1820+070 \citep{Munoz2019}, and GRS~1716--249 \citep{Cuneo2020}. However, these sources either follow a standard outburst evolution (e.g., V404~Cyg; \citealt{Munoz2016}), or the nature of the low inclination affects the visibility of the X-ray disk winds during the soft state (e.g., MAXI~J1820+070; \citealt{Munoz2019}).


We have presented X-ray spectroscopic observations sampling the entire 2021 outburst of the BH transient \src. Our nine-epoch data set reveals the presence of blue-shifted absorption in the intermediate state and soft state, while no absorption line is detected in the hard state. \cite{Mata2022} presented optical spectra taken across the entire outburst with the Gran Telescopio Canarias and Very Large Telescope. The optical spectra indicates the presence of an optical wind in the hard state. The remaining spectra, obtained during the transition to the soft state and the subsequent soft state, are not characterized by optical wind features. Absorption features are also detected in the NIR spectrum of the intermediate state \citep{Mata2022}, but the evolution of the NIR wind is lack because of the short of observation in hard and soft state.

Our study of the X-ray disk winds completes the evolution of disk winds of the source across the entire outburst, as shown in Figure~\ref{wind_tran}. The emergence of X-ray disk winds follows the disappearance of optical wind, accompanying the state transition from hard to soft state. \src\ becomes the first source in which the visibility of the wind across the outburst and as a function of the spectral band is demonstrated from a purely observational point of view. This observational evidence suggests compelling evidence for the optical wind to X-ray wind transition in the \src, and offers fundamental new insights into the long-term coupling of accretion disks and mass outflows around accreting black holes.

Several scenarios that may explain the properties of the X-ray and optical outflows are proposed in \cite{Munoz2022}. In the case of \src, the kinetic properties of the X-ray and optical wind are remarkably consistent with each other, i.e., comparable outflowing velocities, which prefer the presence of a dynamic, multi-phase outflow during the entire outburst of the system. In different accretion state, different-phase outflow play a dominant role. We note that it is difficult to fully rule out other scenarios. The existence of X-ray disk winds in hard state has not been unambiguously studied yet. It remains to be known if the X-ray wind is not launched in the hard state (e.g.,~\citealt{Ueda2010}), or just not detectable because of highly ionized effect (e.g.,~\citealt{Shidatsu2019}) or thermally unstable effect (e.g.,~\citealt{Chakravorty2013}). Towards a unambiguous scenario for the multi-wavelength wind-type outflows, further high-quality observations and theoretical studies are needed.

\section{Conclusion}
\label{conclusion}

We investigated the X-ray disk wind properties of black hole candidate \src\ with \nicer\ data. Spectral analysis on its high-quality $2-10$~keV spectra shows that:

\begin{enumerate}

\item The source exhibited blue-shifted absorption lines in the intermediate and soft state, suggesting the existence of X-ray disk winds during these states. 
\item The thermal nature is suggested for the X-ray disk winds in the source, and the wind carries away less mass than is accreted into the central object.
\item The transition for iron from predominantly H-like to predominantly He-like is detected in the source. Our study suggests that the evolution of the illuminating spectrum and the slow change of the geometric properties of the disk winds together account to this transition. 
\item This study completes the evidence for the evolution of the disk winds, from optical winds to X-ray disk winds in \src.
\end{enumerate}


\section*{Acknowledgements}

This work was supported by the National Natural Science Foundation of China (NSFC), Grant No.~12250610185, 11973019, and 12261131497, and the Natural Science Foundation of Shanghai, Grant No.~22ZR1403400. YZ acknowledges support from the Dutch Research Council (NWO) Rubicon Fellowship, file no.\ 019.231EN.021. The authors wish to thank the anonymous reviewer, whose comments helped clarify and enhance this manuscript. In addition, the author would like to thank Tim Kallman and Yerong Xu for useful guidance on the use of \texttt{XSTAR}.







\begin{table*}[]
\centering
\begin{tabular}{lcccclccc}
\hline\hline

Epoch \hspace{0.2cm} & \hspace{0.2cm} Obs.~ ID  \hspace{0.1cm} & \hspace{0.1cm}  Start data \hspace{0.1cm} & \hspace{0.1cm}  Exposure (s) \hspace{0.1cm} & \hspace{0.1cm} &Epoch \hspace{0.2cm} & \hspace{0.2cm} Obs.~ ID  \hspace{0.1cm} & \hspace{0.1cm}  Start data \hspace{0.1cm} & \hspace{0.1cm}  Exposure (s) \hspace{0.1cm}  \\ \hline
\multirow{4}{*}{Epoch~1}& 4202130101 & 2021-05-02 & 516 & & \multirow{8}{*}{Epoch~8} & 4675020124  & 2021-06-17 & 2876 \\
                      & 4202130102 & 2021-05-03 & 535  & &  & 4202130122  & 2021-06-18 & 2976 \\
                      & 4202130103 & 2021-05-04 & 566  & &   & 4675020125  & 2021-06-18 & 2583 \\
                      & 4202130104 & 2021-05-05 & 4039  &  &  & 4675020126  & 2021-06-19 & 3679 \\
\cline{1-4}
\multirow{3}{*}{Epoch~2} & 4202130105  & 2021-05-18 & 2680  & & & 4675020127  & 2021-06-20 & 2177 \\
                       & 4202130106  & 2021-05-19 & 12211 & & & 4675020128  & 2021-06-21 & 3514 \\
                       & 4202130107  & 2021-05-20 & 4631 & & & 4202130123  & 2021-06-21 & 1092 \\
\cline{1-4}
\multirow{7}{*}{Epoch~3} & 4675020101  & 2021-05-21 & 5814 & & & 4675020129  & 2021-06-22 & 2280 \\
\cline{6-9}
                       & 4202130108  & 2021-05-21 & 3809 & &  \multirow{12}{*}{Epoch~9} & 4675020130  & 2021-06-23 & 1352 \\
                       & 4675020102  & 2021-05-22 & 3011 & & & 4675020132  & 2021-06-26 & 1631 \\
                       & 4202130109  & 2021-05-22 & 3020 & & & 4675020133  & 2021-06-27 & 851 \\
                       & 4675020103  & 2021-05-23 & 5372 & & & 4675020134  & 2021-06-28 & 1450 \\
                       & 4202130110  & 2021-05-23 & 7058 & & & 4675020135  & 2021-06-29 & 1928 \\
                       & 4202130111  & 2021-05-23 & 2456 & & & 4675020137  & 2021-07-01 & 2887 \\
\cline{1-4}
\multirow{7}{*}{Epoch~4} & 4675020104  & 2021-05-24 & 2671 & & & 4675020138  & 2021-07-02 & 1567 \\
                       & 4202130112  & 2021-05-25 & 3683 & & & 4675020139  & 2021-07-03 & 2276 \\
                       & 4675020105  & 2021-05-25 & 2535 & & & 4675020140  & 2021-07-04 & 1906 \\
                       & 4675020106  & 2021-05-26 & 1867 & & & 4675020141  & 2021-07-05 & 2846 \\
                       & 4642010101  & 2021-05-26 & 4318 & & & 4675020142  & 2021-07-06 & 1913 \\
\cline{6-9}
                       & 4642010102  & 2021-05-27 & 768 & &  \multirow{10}{*}{Epoch~10} & 4675020143  & 2021-07-07 & 2562 \\
\cline{1-4}
\multirow{7}{*}{Epoch~5} & 4202130114  & 2021-05-30 & 1749 & & & 4675020144  & 2021-07-08 & 805 \\
                       & 4675020108  & 2021-05-30 & 3574 & & & 4675020145  & 2021-07-09 & 2472 \\
                       & 4675020109  & 2021-05-31 & 5859 & & & 4675020146  & 2021-07-10 & 224 \\
                       & 4202130115  & 2021-05-31 & 3912 & & & 4675020147  & 2021-07-11 & 4275 \\
                       & 4675020110  & 2021-05-31 & 4791 & & & 4675020148  & 2021-07-12 & 1789 \\
                       & 4202130116  & 2021-06-01 & 2341 & & & 4675020149  & 2021-07-24  & 3063 \\
                       & 4675020111  & 2021-06-02 & 5120 & & & 4675020150  & 2021-07-25  & 1731 \\
\cline{1-4}
\multirow{6}{*}{Epoch~6} & 4202130118  & 2021-06-03 & 7394 & & & 4675020151  & 2021-07-26  & 2119 \\
                       & 4675020112  & 2021-06-03 & 4631 & & & 4675020152  & 2021-07-27  & 2350 \\
\cline{6-9} 
                       & 4202130119  & 2021-06-04 & 4872 & & & & &\\
                       & 4675020113  & 2021-06-04 & 4543 & & & & &\\
                       & 4202130120  & 2021-06-05 & 479 & & & & &\\
                       & 4675020114  & 2021-06-06 & 6107 & & & & &\\
\cline{1-4}
\multirow{9}{*}{Epoch~7} & 4675020116  & 2021-06-08 & 3862  & & & & &\\
                       & 4202130121  & 2021-06-09 & 1392 & & & & &\\
                       & 4675020117  & 2021-06-10 & 829 & & & & &\\
                       & 4675020118  & 2021-06-11 & 2662 & & & & &\\
                       & 4675020119  & 2021-06-12 & 3222 & & & & &\\
                       & 4675020120  & 2021-06-13 & 1860 & & & & &\\
                       & 4675020121  & 2021-06-14 & 2874 & & & & &\\
                       & 4675020122  & 2021-06-15 & 1608 & & & & &\\
                       & 4675020123  & 2021-06-16 & 2081 & & & & &\\
\hline\hline
\end{tabular}
\vspace{0.3cm}
\caption{\rm \nicer\ observations of \src\ analyzed in this work.
\label{nicer_obs}}
\end{table*}


\begin{table*}[]
\centering
\hspace{-1.5cm}
\renewcommand\arraystretch{1.1}
\begin{tabular}{lcccccc}
\hline\hline
&  & Epoch~1& Epoch~2 & Epoch~3 & Epoch~4  & Epoch~5 \\ \hline
\texttt{Tbabs}      & $N_{\rm H}$ [$10^{22}$~cm$^{-2}$] & ${0.15}^{*}$         & $< 0.42$   & $< 0.44$  & $0.17_{-0.06}^{+0.18}$ & $0.31_{-0.07}^{+0.04}$   \\
\texttt{xstar}        & $N_{\rm H}$ [10$^{22}$~cm$^{-2}$] & --                          & $2.3_{-0.8}^{+2.9}$       & $1.9_{-0.9}^{+8}$   & $7_{-5}^{+7}$          & $5.4_{-1.5}^{+8}$     \\
                           & rlogxi                                                  & --                         & $4.31_{-0.17}^{+0.4}$  & $4.3_{-0.4}^{+4}$   & $4.8_{-0.5}^{+0.3}$        & $4.84_{-0.12}^{+0.5}$   \\
                           & $z$                                                     & --                         & $> -0.005$ & $> -0.009$ & $> -0.006$ & $> -0.003$   \\
\texttt{diskbb}      & $T_{\rm in}$ [$keV$]                         & $0.28_{-0.06}^{+0.05}$ & $0.64_{-0.04}^{+0.11}$ & $0.62_{-0.05}^{+0.03}$ & $0.81_{-0.11}^{+0.07}$ & $0.91_{-0.04}^{+0.04}$  \\                      
\texttt{nthcomp}   & $\Gamma$                                        & $1.595_{-0.017}^{+0.018}$ & $3.294_{-0.014}^{+0.011}$ & $2.703_{-0.026}^{+0.03}$ & $2.86_{-0.18}^{+0.16}$ & $2.9_{-0.8}^{+0.6}$       \\
\texttt{gauss}$^1$      & $E_{\rm line}$                                   & $2.35_{-0.08}^{+0.08}$ & $2.403_{-0.04}^{+0.029}$ & $2.397_{-0.019}^{+0.019}$ & $2.41_{-0.05}^{+0.05}$ & $2.41_{-0.04}^{+0.04}$   \\                       
                            & $\sigma$  [$keV$]                             & ${0.1}^{*}$          & $0.05^{*}$    &  $0.061_{-0.028}^{+0.029}$         & $0.07^{*}$    & $0.03^{*}$           \\
                            & norm  $\times 10^{-4}$                     & $5_{-3}^{+3}$      & $17_{-9}^{+24}$        & $20_{-8}^{+14}$      & $13_{-4}^{+16}$       & $6.2_{-4}^{+7}$              \\     
\texttt{gauss}$^2$       & $E_{\rm line}$                                  & $2.67_{-0.07}^{+0.06}$ & $2.695_{-0.024}^{+0.025}$ & $2.67_{-0.05}^{+0.05}$ & $2.702_{-0.03}^{+0.026}$ & $2.70_{-0.04}^{+0.04}$            \\               
                            & $\sigma$  [$keV$]                           & ${0.05}^{*}$         & $< 0.05$ & ${0.1}^{*}$  & $< 0.06$  & $< 0.09$             \\
                            & norm $\times 10^{-4}$                    & $2.1_{-1.5}^{+1.6}$ & $7.9_{-4}^{+5}$         &  $11_{-5}^{+5}$                  & $6_{-3}^{+3}$            & $8_{-4}^{+4}$           \\               
\texttt{gauss}$^3$       & $E_{\rm line}$                                 & $< 2.07$   & $< 2.07$  & $< 2.06$  & $< 2.07$  & $2.0^{*}$          \\
                            & $\sigma$  [$keV$]                           & $0.10^{*}$                  & $> 0.04$  & $> 0.05$  & $0.06^{*}$               & $0.01^{*}$           \\
                            & norm  $\times 10^{-4}$                   & $< 8$             & $42_{-27}^{+60}$       & $49_{-28}^{+50}$      & $< 42$       & $< 26$               \\
\texttt{gauss}$^4$         & $\sigma$  [$keV$]                           & $0.94_{-0.14}^{+0.16}$ & $0.94^{*}$          & $0.84_{-0.17}^{+0.16}$       & $0.96^{*}$                 & $1.01_{-0.3}^{+0.23}$     \\                  
\hline
                           & $\log F_{\rm bb}$                             & $-9.65_{-0.06}^{+0.16}$ & $-8.56_{-0.28}^{+0.3}$ & $-8.473_{-0.04}^{+0.027}$ & $-8.20_{-0.08}^{+0.04}$ & $-8.085_{-0.025}^{+0.018}$   \\
                           & $\log F_{\rm nth}$                            & $-8.558_{-0.008}^{+0.007}$ & $-7.959_{-0.12}^{+0.024}$ & $-8.251_{-0.026}^{+0.04}$ & $-8.25_{-0.06}^{+0.15}$ & $-8.93_{-0.22}^{+0.17}$      \\
                            & $\log F_{\rm gau} (\rm reflction)$    & $-10.20_{-0.10}^{+0.10}$ & $< -12.4$  & $-10.58_{-0.19}^{+0.13}$ & $-10.64_{-0.9}^{+0.23}$ & $-10.8_{-0.7}^{+0.4}$        \\             
\hline                                                                                
$\chi^2/\nu$       &                                                          & 758.46/789                        & 604.44/780                            & 664.50/781                                   & 608.47/780                           & 665.10/780        \\
\hline \hline  
&  & Epoch~6 & Epoch~7 & Epoch~8 & Epoch~9  & Epoch~10 \\ \hline
\texttt{Tbabs}      & $N_{\rm H}$ [$10^{22}$~cm$^{-2}$] & $0.27_{-0.04}^{+0.04}$ & $0.27_{-0.04}^{+0.03}$ & $0.22_{-0.04}^{+0.04}$ & $0.172_{-0.021}^{+0.04}$  & $0.15_{-0.06}^{+0.06}$   \\
\texttt{xstar}        & $N_{\rm H}$ [10$^{22}$~cm$^{-2}$] & $8.3_{-2.1}^{+3}$          & $8.0_{-1.5}^{+2.3}$ & $9.5_{-2.5}^{+2.9}$                          & $7.6_{-2.5}^{+4}$     & $9_{-4}^{+5}$  \\
                           & rlogxi                                                  & $4.96_{-0.14}^{+0.17}$ & $4.82_{-0.11}^{+0.16}$ & $5.07\pm 0.18$ & $4.96_{-0.17}^{+0.26}$       & $5.0\pm 0.3$  \\
                           & $z$                                                     & $> -0.0008$ & $> -0.0020$ & $> -0.0011$ & $> -0.0013$   & $> -0.003$          \\
\texttt{diskbb}      & $T_{\rm in}$ [$keV$]                         & $0.880_{-0.004}^{+0.006}$ & $0.852_{-0.004}^{+0.011}$ & $0.830_{-0.005}^{+0.006}$ & $0.768_{-0.006}^{+0.008}$      & $0.703_{-0.006}^{+0.007}$        \\                  
\texttt{nthcomp}   & $\Gamma$                                        & ${4.0}^{*}$          & $> 3.1$    & ${4.0}^{*}$          & ${4.0}^{*}$           & $> 3.6$         \\
\texttt{gauss}$^1$       & $E_{\rm line}$                                   & $2.38_{-0.25}^{+0.06}$ & $2.399_{-0.025}^{+0.029}$ & $2.382_{-0.019}^{+0.02}$ & $2.40_{-0.05}^{+0.07}$       & $2.384_{-0.024}^{+0.024}$   \\                       
                            & $\sigma$  [$keV$]                             & $0.031^{*}$               & $< 0.06$  & $< 0.05$  & $> 0.01$             & $<0.05$                 \\
                            & norm  $\times 10^{-4}$                     & $7.6_{-6}^{+20}$       & $4.8_{-2.3}^{+2.4}$ & $6.2_{-2.0}^{+2.9}$ & $5.1_{-2.5}^{+4}$          & $2.6_{-1.2}^{+1.3}$               \\     
\texttt{gauss}$^2$      & $E_{\rm line}$                                  & $2.67_{-0.06}^{+0.05}$ & $2.656_{-0.029}^{+0.028}$ & $2.64_{-0.04}^{+0.03}$ & $2.66_{-0.03}^{+0.03}$        & $2.637_{-0.026}^{+0.028}$                  \\         
                            & $\sigma$  [$keV$]                            & $0.01^{*}$    & $> 0.02$ & $> 0.04$ & $< 0.09$       & $0.06_{-0.03}^{+0.04}$                  \\
                            & norm $\times 10^{-4}$                     & $2.8_{-2.1}^{+3}$      & $9_{-3}^{+4}$ & $7.8_{-2.7}^{+3}$ & $4.9_{-2.1}^{+2.3}$          & $4.4_{-1.5}^{+1.7}$             \\               
\texttt{gauss}$^3$      & $E_{\rm line}$                                  & $2.22_{-0.04}^{+0.13}$ & $< 2.04$ & $< 2.02$ & $< 2.02$         & $< 2.02$                  \\
                            & $\sigma$  [$keV$]                           & $> 0.06$  & $< 0.07$  & $< 0.06$ & $< 0.05$            & $< 0.04$                      \\
                            & norm  $\times 10^{-4}$                    & $-34_{-18}^{+14}$    & $9.9_{-5}^{+12}$ & $12_{-5}^{+11}$  & $6_{-4}^{+4}$        & $7.0_{-2.9}^{+3}$                     \\
\texttt{gauss}$^4$         & $\sigma$  [$keV$]                           & $1.0_{-0.6}^{+0.7}$  & $0.80_{-0.17}^{+0.21}$ & $1.19_{-0.23}^{+0.27}$ & $1.06_{-0.24}^{+0.27}$            & $1.07_{-0.21}^{+0.22}$                  \\                  
\hline
                           & $\log F_{\rm bb}$                             & $-8.1368_{-0.0023}^{+0.003}$ & $-8.2195_{-0.0025}^{+0.006}$ & $-8.296_{-0.003}^{+0.003}$ & $-8.432_{-0.003}^{+0.003}$  & $-8.602_{-0.004}^{+0.004}$           \\
                           & $\log F_{\rm nth}$                            & $-9.17_{-0.04}^{+0.02}$ & $-9.58_{-0.23}^{+0.04}$ & $-10.07_{-0.4}^{+0.14}$ & $-10.01_{-0.28}^{+0.12}$        & $-10.14_{-0.19}^{+0.11}$           \\
                            & $\log F_{\rm gau} (\rm reflction)$    & $-11.4_{-0.7}^{+0.6}$ & $-11.06_{-0.18}^{+0.16}$ & $-10.92_{-0.22}^{+0.21}$ & $-11.10_{-0.26}^{+0.25}$         & $-11.27_{-0.24}^{+0.20}$              \\     
\hline                                                                                        
$\chi^2/\nu$       &                                                         & 655.52/781                                 & 622.24/768                                & 690.65/747                                     & 757.68/781                       & 551.47/656                    \\
\hline \hline   
\end{tabular}
\vspace{0.3cm}
\caption{Best-fit results with \texttt{XSTAR} model. \texttt{gauss}$^1$, \texttt{gauss}$^2$, and \texttt{gauss}$^3$ are for the calibration issues in soft energy band, in which a upper limit of 0.1~keV is considered for $\sigma$ parameters. \texttt{gauss}$^4$ with a central energy fixed at 6.4~keV account for the reflection feature (broad Fe K$\alpha$ line). The flux (1.0--10.0~keV) of the each component are presented in units of erg~s$^{-1}$~cm$^{-2}$. All uncertainties are quoted at the 90\% confidence level.
\label{bestfit}}
\end{table*}

\bibliography{sample631}


\end{document}